\pdfoutput=1

\documentclass[11pt]{article}

\usepackage{acl}

\usepackage[algo2e]{algorithm2e} 
\usepackage{algorithm, algorithmic}

\usepackage{times}
\usepackage{arydshln}
\usepackage{ulem}
\usepackage{latexsym}
\usepackage{booktabs}
\usepackage{amsmath,amsfonts}
\usepackage{multirow}
\usepackage{arydshln}
\usepackage{colortbl}
\usepackage{booktabs}
\definecolor{darkergreen}{rgb}{0.0, 0.5, 0.0}
\definecolor{darkerred}{rgb}{0.5, 0.0, 0.0}
\usepackage[framemethod=TikZ]{mdframed}
\usepackage[T1]{fontenc}
\definecolor{revised}{RGB}{0, 0, 255}

\usepackage[utf8]{inputenc}

\usepackage{microtype}

\newcommand\blfootnote[1]{%
  \begingroup
  \renewcommand\thefootnote{}\footnote{#1}%
  \addtocounter{footnote}{-1}%
  \endgroup
}

\title{ProSE: Diffusion Priors for Speech Enhancement}

\author{
    Sonal Kumar$^{*}$,
    Sreyan Ghosh$^{*}$,
    Utkarsh Tyagi,
    Anton Jeran Ratnarajah,\\
    \bf Chandra Kiran Reddy Evuru,
    \bf  Ramani Duraiswami$^{\dagger}$,
    \bf Dinesh Manocha$^{\dagger}$ \\
    University of Maryland, College Park, USA \\
    \texttt{\{sonalkum,sreyang,utkarsht\}@umd.edu} \\ 
}

\begin{document}
\maketitle
\begin{abstract}
Speech enhancement (SE) is the foundational task of enhancing the clarity and quality of speech in the presence of non-stationary additive noise. While deterministic deep learning models have been commonly employed for SE, recent research indicates that generative models, such as denoising diffusion probabilistic models (DDPMs), have shown promise. However, unlike speech generation, SE has a strong constraint in generating results in accordance with the underlying ground-truth signal. Additionally, for a wide variety of applications, SE systems need to be employed in real-time, and traditional diffusion models (DMs) requiring many iterations of a large model during inference are inefficient. To address these issues, we propose ProSE (diffusion-based \textbf{\underline{Pr}}i\textbf{\underline{o}}rs for \textbf{\underline{SE}}), a novel methodology based on an alternative framework for applying diffusion models to SE. Specifically, we first apply DDPMs to generate priors in a latent space due to their powerful distribution mapping capabilities. The priors are then integrated into a transformer-based regression model for SE. The priors guide the regression model in the enhancement process. Since the diffusion process is applied to a compact latent space, the diffusion model takes fewer iterations than the traditional DM to obtain accurate estimations. Additionally, using a regression model for SE avoids the distortion issue caused by misaligned details generated by DMs. Our experiments show that ProSE achieves state-of-the-art performance on benchmark datasets with fewer computational costs. Our code is available on GitHub\footnote{\url{https://github.com/sonalkum/ProSE}}.
\blfootnote{${^*}$Equal contribution.$^\dagger$Equal Advising.}
\end{abstract}

\setlength{\abovedisplayskip}{6pt}
\setlength{\belowdisplayskip}{6pt}

\section{Introduction}

Speech enhancement (SE) is a task that focuses on enhancing speech intelligibility and quality, especially when speech is degraded by non-stationary additive noise. SE has important applications in various areas, including telecommunications~\cite{gay2012acoustic}, medicine~\cite{van2009speech}, and everyday communication~\cite{tashev2011recent}. Traditionally, deep learning models have been employed to establish a deterministic mapping from noisy to clean speech. Although deterministic models have been considered superior in SE, recent developments in generative models have shown promise and narrowed the performance gap~\cite{lu2021study,richter2023speech,9746901}.


Denoising diffusion probabilistic models (DDPMs) have emerged as a powerful generative approach for realistic speech synthesis~\cite{kong2021diffwave,zhang2021restoring,shen2023naturalspeech}. Specifically, DDPMs are designed to denoise data iteratively by inverting the diffusion process. These models demonstrate that structured probabilistic diffusion approaches can effectively transform randomly sampled Gaussian noise into complex target distributions, such as realistic speech or associated latent distributions, while avoiding the mode collapse and training instabilities commonly seen in GANs~\cite{creswell2018generative,donahue2018exploring}. However, there are two major problems: \textbf{(1)} As a class of likelihood-based models, DDPMs require many steps in large denoising models to generate precise details of the data distributions, which requires massive computational resources. Unlike the speech synthesis tasks that generate each detail from scratch, SE tasks are only required to remove additive noise. Therefore, adopting the same extensive iterative process used for speech synthesis in SE leads to inefficiencies and excessive computational costs. \textbf{(2)} The multi-step denoising process in DDPMs can cause misalignment between the original clean speech and the enhanced output, potentially introducing artifacts that degrade the quality of the enhanced speech~\cite{tai2023dose}.
\vspace{1cm}

{\noindent \textbf{Main Contributions.}} To address the challenges outlined above while efficiently leveraging the powerful distribution mapping abilities of DDPMs, we propose ProSE, an alternative and novel framework for applying DDPMs to SE. Specifically, we apply DDPMs to generate priors in a latent space. These priors are then integrated into a Transformer-based regression model. Our regression model is designed in the shape of a U-Net~\cite{li2023dcht,ronneberger2015u}, and we integrate the DDPM-generated priors into the model using cross-attention to guide the model for SE. ProSE is trained in a unique 2-stage manner: \textbf{(1)} In the first stage, we learn a latent encoder that can compress clean speech into a highly compact latent space. We employ a convolution-based latent encoder to first generate a latent space for clean speech. This then acts as the prior and is conditioned on the regression model to generate clean speech from its noisy version. In this step, both the latent encoder and the regression model are jointly trained in an end-to-end (E2E) fashion. \textbf{(2)} In the second stage, we learn a latent diffusion model (LDM) that can generate the prior from Gaussian noise. For training, the latent learned in stage 1 acts as the starting point of the forward diffusion process, and the LDM-generated latent acts as the prior for conditioning the regression model. As in stage 1, we train the LDM and the regression model in a joint fashion. This mitigates error propagation from one component to another. To summarize, our main contributions are as follows:
\begin{enumerate}
\vspace{-2mm}
\setlength\parskip{0em}
    \item We propose ProSE, a novel methodology to leverage the powerful distribution mapping abilities of DDPMs for SE. ProSE employs LDMs to first generate a prior in the latent space that is then conditioned on a Transformer-based regression model to guide the model for SE. ProSE overcomes existing problems with employing LDMs in SE, including the requirement of a large number of inference steps and their tendency to produce undesired artifacts not present in the original clean speech. ProSE is trained in a unique 2-stage fashion wherein, at each stage, all components of the model are trained end-to-end.
    \item Experiment on benchmark datasets demonstrate that our method surpasses recent diffusion-based models for SE in terms of both accuracy and efficiency, including in mismatched scenarios.
\end{enumerate}

\section{Related Work}
\label{sec:related}

SE methods that employ DDPMs have been extensively studied in literature. There are 2 dominant methods that the community has been focused on:
\vspace{0.5mm}

\noindent \textbf{(1) Condition-injecting Strategies.} Given clean data, diffusion models generally add Gaussian noise to the data in multiple steps (according to a noise schedule) in the forward diffusion process to later estimate the noise in the reverse diffusion process. Though applicable to diverse speech generation tasks, this is not applicable to SE as additive noise is not always Gaussian in nature. Thus, to include actual noisy speech in the process, researchers have found ways to linearly interpolate between clean and noisy speech~\cite{lu2021study,lu2022conditional} or to perform this transformation using the drift term of a stochastic differential equation (SDE)~\cite{richter2023speech,welker2022speech}.
\vspace{0.5mm}

\noindent \textbf{(2)} \textbf{Auxiliary Conditioning Strategies.} These methods devise auxiliary conditions to guide the LDM to generate clean speech from Gaussian noise with diffusion models~\cite{zhang2021restoring,tai2023dose,serra2022universal}. Leveraging conditioning is generally challenging~\cite{tai2023revisiting}, and thus, these models need specific architectures.

ProSE is like type 2 in its motivation; however, it employs an alternative framework. Our objective is not to devise a good auxiliary condition to guide DDPMs for SE but rather to generate a good latent prior with DDPMs to guide a self-attention-based regression model for SE.

\begin{figure*}[t]
    \centering
\includegraphics[width=\textwidth]{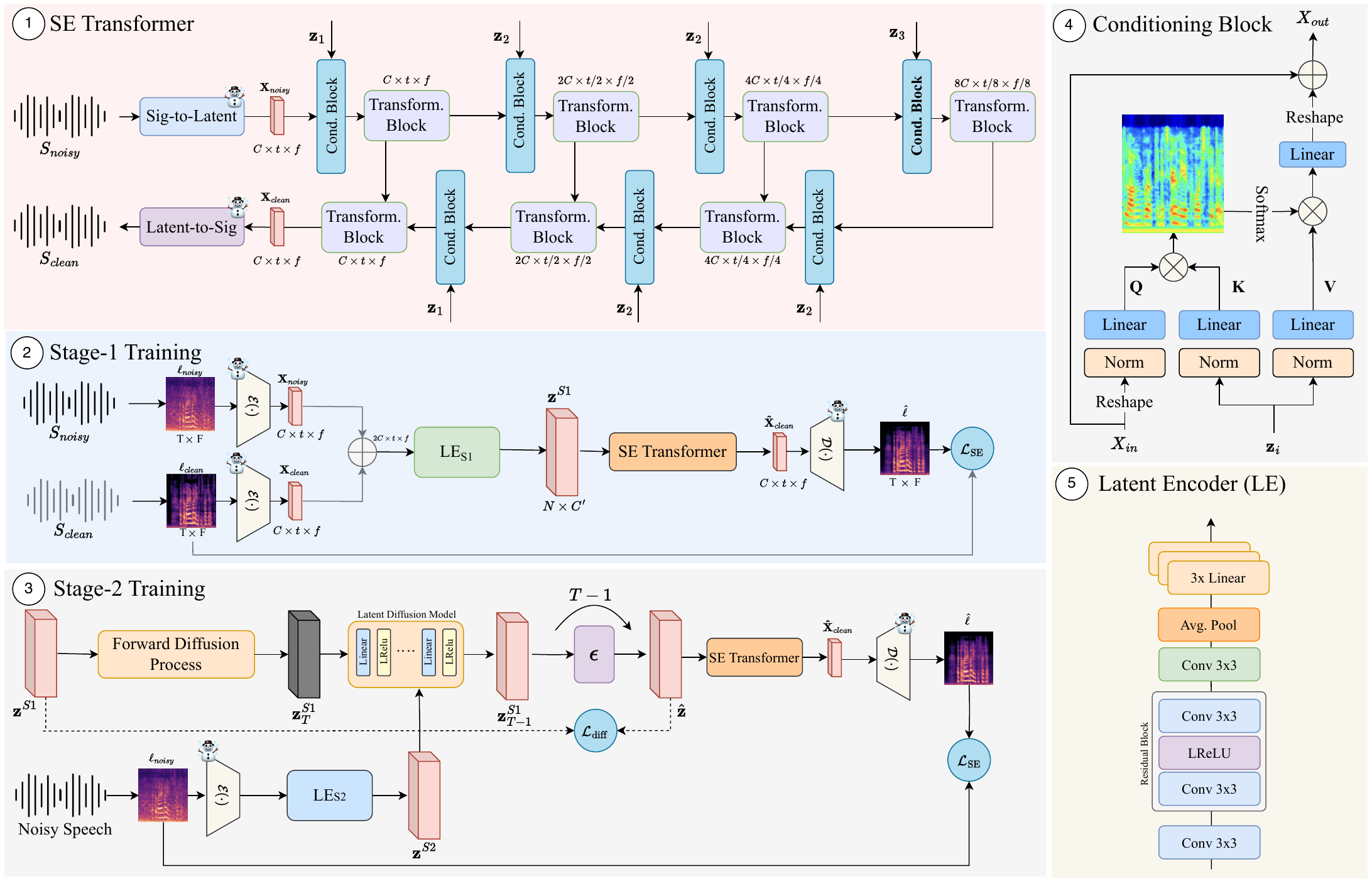}
    \caption{\small Illustration of \textbf{ProSE.} \textcircled{\raisebox{-0.9pt}{1}} The SE Transformer adopts a hierarchical encoder-decoder architecture conditioned on priors to guide the SE process. \textcircled{\raisebox{-0.9pt}{2}} In Training Stage 1, we aim to learn a latent encoder $\operatorname{LE_{S1}}$ that can compress the clean speech to  $\mathbf{z}^{\operatorname{S1}}$. We first obtain latent encodings of the clean and noisy speech $\mathbf{X}_{clean}$ and $\mathbf{X}_{noisy}$ after mel-spectrogram conversion and VAE encoding. Next, we concatenate both along the channel dimension before feeding it into $\operatorname{LE_{S1}}$. $\mathbf{z}^{\operatorname{S1}}$ is then integrated with the SE transformer as the prior ($\mathbf{z}^{\operatorname{S1}}$=$\mathbf{z}_{1}$) and $\operatorname{LE_{S1}}$ and the SE transformer is trained E2E. \textcircled{\raisebox{-0.9pt}{3}} For Training Stage 2, we now employ $\mathbf{z}^{\operatorname{S1}}$ from $\operatorname{LE_{S1}}$ as the starting point of the forward diffusion process to train the LDM. The LDM iteratively removes noise from $\mathbf{z}^{\operatorname{S1}}_T$, to obtain $\hat{\mathbf{z}}$, and is conditioned on $\mathbf{z}^{\operatorname{S2}}$, a compact latent of the noisy speech obtained from another latent encoder $\operatorname{LE_{S2}}$. For Stage 2, the LDM and the Transformer are trained E2E, where $\hat{\mathbf{z}}$=$\mathbf{z}_{1}$ and $\hat{\mathbf{z}}$ is further downsampled. For inference, we drop $\operatorname{LE_{S1}}$ and start the reverse process from randomly sampled Gaussian noise.}
    \label{fig:main_diag}
    \vspace{-2mm}
\end{figure*}

\section{Methodology}
\label{sec:method}

{\noindent \textbf{Primary Objective.}} Our objective is to integrate generative DDPMs with deterministic regression-based models. Through this, we aim to leverage each of their strengths, mitigating the limitations inherent in DDPMs while harnessing their powerful distribution mapping capabilities. Specifically, we leverage LDMs to generate a prior, which is then hierarchically integrated into a Transformer-based regression model. The prior acts as conditioning and guides the Transformer model for SE. The following subsections will describe the ProSE architecture followed by the 2-step training mechanism. 


{\noindent \textbf{Background on Diffusion models.}} Diffusion models consist of two main processes: a forward process and a reverse process. Given a data point $x_0$ with probability distribution $p(x_0)$, the forward diffusion process, 
gradually adds Gaussian noise to $x_0$ according to a pre-set variance schedule $\beta_1, \cdots, \beta_T$ and degrades the structure of the data. At the time step $t$, the latent variable ${x_t}$ is only determined by $x_{t-1}$ due to its  discrete-time Markov process nature, and can be expressed as:
\begin{equation}
    p(x_t \mid x_{t-1}) = \mathcal{N}(x_t; \sqrt{1 - \beta_t} x_{t-1}, \beta_t I), 
\end{equation}
As $t$ increases over several diffusion steps, $p(x_T)$ approaches a unit spherical Gaussian distribution. The marginal distribution of $x_t$ at any given step can be expressed analytically as:
\begin{equation}
    p(x_t \mid x_0) = \mathcal{N}(x_t; \sqrt{\alpha_t} x_0, (1 - \alpha_t) I), 
\end{equation}
where $\alpha_t = \prod_{s=1}^t (1 - \beta_s)$. The reverse process aims to reconstruct the original data from the noise-corrupted version by learning a series of conditional distributions. The transition from $x_t$ to $x_{t-1}$ is modeled as:
\begin{equation}
p_\theta(x_{t-1} \mid x_t) = \mathcal{N}(x_{t-1}; \mu_\theta^{t-1},  \sigma_\theta^{t-1}),
\end{equation}
\begin{equation}
\mu_\theta^{t-1} = \frac{x_t - \beta_t \epsilon_\theta(x_t, t)}{\sqrt{1-\beta_t}},
\end{equation}
\begin{equation}
\sigma_\theta^{t-1^2} =\frac{1-\bar{\alpha}_{t-1}}{1-\bar{\alpha}_t} \cdot \beta_t,
\end{equation}where $\alpha_t=1-\beta_t, \bar{\alpha}_t=\prod_{i=1}^t \alpha_i$, $\theta$ represents the learnable parameters, $\mu_\theta^{t-1}$ is the mean estimate, $\sigma_\theta^{t-1^2}$ is the standard deviation estimate, and $\epsilon_\theta(x_t, t)$ is the noise estimated by the neural network. The reverse process estimates the data distribution $p(x_0)$ by integrating over all possible paths:
\begin{equation}
    p_\theta(x_0) = \int p_\theta(x_T) \prod_{t=1}^T p_\theta(x_{t-1} \mid x_t) \, dx_1:T
\end{equation}
where $p_\theta(x_T) = \mathcal{N}(x_T; 0, I)$. 

\subsection{ProSE Architecture}
\label{sec:arch}

{\noindent \textbf{Overview.}} The overall architecture of ProSE is illustrated in Fig.~\ref{fig:main_diag}. Given a noisy speech signal, we first transform it into a mel-spectrogram representation, which is then compressed using a Variational Auto Encoder (VAE). The compressed output is fed into our main speech enhancement (SE) modules, comprising a latent encoder (LE), an LDM, and a Transformer. The flow of information differs in different stages of training, and we elaborate on this in a later sections. The Transformer generates a latent representation that is sent to the VAE decoder to reconstruct a mel-spectrogram. Subsequently, we utilize a HiFi-GAN Vocoder to convert the mel-spectrogram back into a clean speech signal. We will now describe each component in detail.

{\noindent \textbf{Mel-Spectrogram Conversion (Signal-to-Mel).}} Given a speech signal $\mathbf{S}$, we first convert the signal into a mel-spectrogram  $\ell$ for further processing. For speech or audio synthesis, including SE, diffusion models have been studied for both mel-spectrogram generation~\cite{qiang2024high,chen2022resgrad} and waveform generation~\cite{lam2022bddm,lee2021priorgrad}. We choose the former due to its ease of integration with existing frameworks and better performance.
\vspace{0.5mm}

{\noindent \textbf{VAE.}} Given the mel-spectrogram $\ell \in \mathbb{R}^{\text{T} \times \text{F}}$, we use a VAE encoder $\mathcal{E}(\cdot)$ to encode it to a latent space $\mathbf{X} \in \mathbb{R}^{C \times t \times f}$ where $t = \frac{\text{T}}{r}$, $f = \frac{\text{F}}{r}$ and $r$ is the compression level of the latent space. The SE transformer outputs $\hat{\mathbb{R}} \in \mathbb{R}^{C \times t \times f}$, and we use the VAE decoder $\mathcal{D}(\cdot)$ to transform it back to $\hat{\ell} \in \mathbb{R}^{\text{T} \times \text{F}}$. $\hat{\ell}$ can now be passed to a vocoder to obtain the predicted clean speech signal $\hat{\mathbf{S}}_{clean}$.

Our VAE is composed of an encoder and a decoder with stacked convolutional modules. During training, we adopt a reconstruction loss, an adversarial loss, and a Gaussian constraint loss. Architecture and training methods are detailed in Appendix C. In the sampling process, the decoder is used to reconstruct the mel-spectrogram. 

{\noindent \textbf{Latent Encoder (LE).}} Given the latent $\mathbf{X}$, our objective is to compress it to a highly compact latent space $\mathbf{z} \in \mathbb{R}^{N \times C^{\prime}}$. $N$ and $C^{\prime}$ are the token number and channel dimensions, respectively, and $N$ is sufficiently reduced from $\mathbf{X}$ or $\ell$. We employ $\mathbf{z}$ as conditioning to the SE Transformer and the LDM, which we describe later.

We illustrate the latent encoder archtecture in Fig.~\ref{fig:main_diag}. The layer begins with a convolution layer followed by the core, which features $L$ residual blocks, each comprising two convolutional layers, followed by LeakyReLU activation and an additional convolutional layer that feeds into the residual connection. Post-residual processing, an average pooling layer reduces spatial dimensions, succeeded by a sequence of linear transformations and reshaping operations to produce the final output.
\vspace{0.5mm}

{\noindent \textbf{SE Transformer.}} The architecture of the SE Transformer is illustrated in Fig.~\ref{fig:main_diag}. The SE transformer takes as input the latent $\mathbf{X}$ and outputs $\hat{\mathbf{X}}$. The transformer, shaped in the form of a U-Net, consists of 4 encoder blocks and 3 decoder blocks. U-Net architectures have shown promise in SE~\cite{choi2018phase}, speech generation~\cite{li2022unet}, and other audio tasks~\cite{liu2023audioldm}. Each encoder block progressively decreases the time and frequency dimensions of $\mathbf{X}$ while increasing its channels. Each decoder block progressively decreases the time and frequency dimensions of $\mathbf{X}$ while increasing its channels to restore it to its original dimensions. We employ a \textbf{Conditioning Block} in front of each encoder and decoder to integrate $\mathbf{z}$ to guide the enhancement process. 

Specifically, given an intermediate feature $\mathbf{X}_{i n} \in \mathbb{R}^{\hat{t} \times \hat{f} \times \hat{c}}$, we reshaped it as tokens $\mathbf{X}_r \in \mathbb{R}^{(\hat{t} \times \hat{f}) \times \hat{C}}$; where $\hat{t} \times \hat{f}$ is spatial resolution, and $\hat{C}$ denotes channel dimension. Then we linearly project $\mathbf{X}_r$ into the query $\mathbf{Q} \in \mathbb{R}^{(\hat{t} \times \hat{f}) \times \hat{C}}$. Similarly, we project the prior feature $\mathbf{z}_i \in \mathbb{R}^{\hat{N} \times C^{\prime}}$ into the key $\mathbf{K} \in \mathbb{R}^{\hat{N} \times \hat{C}}$ and $\mathbf{V} \in \mathbb{R}^{\hat{N} \times \hat{C}}$ (value). The cross-attention is formulated as:
\begin{equation}
\mathbf{Q}=\mathbf{W}_Q \mathbf{X}_r, \mathbf{K}=\mathbf{W}_K \mathbf{z}_i, \mathbf{V}=\mathbf{W}_V \mathbf{z}_i
\end{equation}
\begin{align}
    \text{Attention}(\mathbf{Q}, \mathbf{K}, \mathbf{V}) 
    &= \operatorname{SoftMax}\left(\frac{\mathbf{Q} \mathbf{K}^T}{\sqrt{\hat{C}}}\right) \cdot \mathbf{V}
\end{align}
where $\mathbf{W}_Q \in \mathbb{R}^{\hat{C} \times \hat{C}}, \mathbf{W}_K \in \mathbb{R}^{C^{\prime} \times \hat{C}}$, and $\mathbf{W}_V \in \mathbb{R}^{C^{\prime} \times \hat{C}}$ represent learnable parameters of linear projections without bias. As vanilla multi-head self-attention~\cite{vaswani2017attention}, we separate channels into multiple "heads" and calculate the attention operations. Finally, we reshape and project the output of cross-attention and add it with $\mathbf{X}_{i n}$ to derive the output feature $\mathbf{X}_{o u t} \in \mathbb{R}^{\hat{t} \times \hat{f} \times \hat{C}}$.

We generate the multi-scale priors \{$\mathbf{z}_1$, $\mathbf{z}_2$, $\mathbf{z}_3$\}, (where $\mathbf{z}_1$=$\mathbf{z}$), by downsampling the prior feature $\mathbf{z}_1$ using a simple \textbf{Downsample} block. The multi-scale prior feature adapts to different intermediate features of different scales for better fusion.
\vspace{0.5mm}

{\noindent \textbf{Latent Diffusion Model (LDM).}} Our LDM is based on conditional DDPMs commonly employed in image and speech generation~\cite{pmlr-v202-liu23f,rombach2021highresolution,ho2020denoising}. Similar to the process discussed in Section~\ref{sec:method}, the LDM involves a forward diffusion process and a reverse denoising process. The exact working and training of the LDM in ProSE is elaborated in Section~\ref{sec:stage_2}.

{\noindent \textbf{HiFi-GAN (Vocoder).}} For vocoder, we employ HiFi-GAN~\cite{kong2020hifi} to generate the predicted clean speech signal $\hat{\mathbf{S}}_{clean}$, from the reconstructed mel-spectrogram $\hat{\ell}$. More details on the vocoder training can be found in Appendix~\ref{sec:hifi}.


\subsection{ProSE Training and Inference}
\label{sec:training}

\subsubsection{Training Stage 1}
\label{sec:stage_1}

For stage 1 of training, we aim to learn a latent encoder that can compress the clean speech signal into the highly compact latent space. We employ this latent encoder in stage 2 to learn the LDM. 

Given clean and noisy speech signals, $\mathbf{S}_{clean}$ and $\mathbf{S}_{noisy}$ respectively, we first obtain its latent encodings $\mathbf{X}_{clean}$ and $\mathbf{X}_{noisy}$ after mel-spectrogram conversion and VAE encoding. Next, we concatenate both of them along the channel dimension and feed them into the latent encoder $\operatorname{LE_{S1}}$ to generate a highly compact prior feature $\mathbf{z}^{\operatorname{S1}} \in \mathbb{R}^{N \times C^{\prime}}$. We then integrate $\mathbf{z}^{\operatorname{{S1}}}$ to the SE transformer hierarchically, as illustrated in Fig.~\ref{fig:main_diag}, to guide the transformer to outputs a latent $\mathbf{\hat{X}}_{clean}$. $\mathbf{\hat{X}}_{clean}$ is then passed through the VAE decoder to finally obtain $\hat{\ell}$ or the predicted clean mel-spectrogram. We jointly train $\operatorname{LE_{S1}}$ and SE transformer end-to-end. The VAE encoder and decoder are kept frozen throughout training. For training, we optimize the $L_1$ loss function as follows:
\begin{equation}
    \mathcal{L}_{\text {SE}}=\left\|\hat{\ell}-\ell\right\|_1,
\end{equation}
where $\hat{\ell}$ is the predicted clean mel-spectrogram  and $\ell$ is the ground truth mel-spectrogram.

\subsubsection{Training Stage 2}
\label{sec:stage_2}

For stage 2 of training ProSE, we aim to learn an LDM to generate the prior feature that enhances the enhancement process of the SE Transformer. 

Given a ground truth clean speech signal, we employ the latent encoder $\operatorname{LE_{S1}}$ from stage one to generate the prior $\mathbf{z}^{\operatorname{S1}}$. This prior $\mathbf{z}^{\operatorname{S1}}$ serves as the initial condition for the \textit{\textbf{forward diffusion process}}, during which Gaussian noise is incrementally added over $T$ iterations as follows:
\begin{table*}[t]
\centering
\resizebox{\textwidth}{!}{
\begin{tabular}{lcccccc||ccccc}
\toprule
\multirow{2}{*}{\textbf{Method}} & \multirow{2}{*}{\textbf{Steps}} & \multicolumn{5}{c||}{\textbf{VBD}}    & \multicolumn{5}{c}{\textbf{TIMIT+MUSAN}}   \\ \cmidrule{3-12} 
&  & \textbf{STOI (\%)}$\uparrow$& \multicolumn{1}{l}{\textbf{PESQ}$\uparrow$}    & \textbf{CSIG}$\uparrow$    & \textbf{CBAK}$\uparrow$    & \textbf{COVL}$\uparrow$    & \textbf{STOI (\%)}$\uparrow$& \textbf{PESQ}$\uparrow$    & \textbf{CSIG}$\uparrow$        & \textbf{CBAK}$\uparrow$    & \textbf{COVL}$\uparrow$    \\ \midrule
\cellcolor{gray!15}Unprocessed  & \cellcolor{gray!15}-                   & \cellcolor{gray!15}92.1    & \cellcolor{gray!15}1.97       & \cellcolor{gray!15}3.35   & \cellcolor{gray!15}2.44   & \cellcolor{gray!15}2.63   & \cellcolor{gray!15}87.27  & \cellcolor{gray!15}1.42 & \cellcolor{gray!15}3.12 & \cellcolor{gray!15}2.43 & \cellcolor{gray!15}2.39   \\ 
\noalign{\vskip 1mm}
\cdashline{1-12}
\noalign{\vskip 1mm}
Diffwave     & 1 step (dis)        & 93.24   & 2.50        & 3.71   & 3.27   & 3.10    & 88.36 & 1.96 & 3.39 & 2.88 & 2.68   \\
DiffuSE      & 6 steps             & 93.47   & 2.37       & 3.70    & 3.03   & 3.03   & 89.44 & 1.79 & 3.32 & 2.83 & 2.62\\
CDiffuSE     & 6 steps             & \underline{93.59}& 2.41       & 3.76   & 3.08   & 3.07   & 90.24 & 1.86 & 3.37 & 2.81 & 2.62   \\
SGMSE        & 50 steps            & 93.20    & 2.34       & 3.70    & 2.90    & 2.99   & 89.02 & 1.74 & 3.29 & 2.73 & 2.59   \\
DR-DiffuSE   & 6 steps             & 92.94   & 2.48       & 3.68   & 3.27   & 3.05   & 87.82 & 1.91 & 3.45 & 2.91 & 2.65   \\
DOSE         & 50 steps            & 93.52   & {\underline{2.54}}& \underline{3.80}& \underline{3.27}& \underline{3.14}&  \underline{90.66} & \underline{2.02} & \underline{3.48} & \underline{2.95} & \underline{2.74}   \\ 
\noalign{\vskip 1mm}
\cdashline{1-12}
\noalign{\vskip 1mm}
SE Transformer \textit{(ours)}      & 50 steps            & $92.98_{\pm{0.03}}$         & {$2.31_{\pm{0.02}}$}         & $3.78_{\pm{0.03}}$         & $3.04_{\pm{0.08}}$         & $2.90_{\pm{0.07}}$         & $88.39_{\pm{0.04}}$ & $1.82_{\pm{0.03}}$ & $3.39_{\pm{0.07}}$ & $2.84_{\pm{0.06}}$ & $2.61_{\pm{0.08}}$         \\
\textbf{ProSE} \textit{(ours)}      & 2 steps             & \cellcolor{magenta!15}$\textbf{94.10}_{\pm{0.07}}$ & {\cellcolor{magenta!15}$\textbf{2.87}_{\pm{0.03}}$} & \cellcolor{magenta!15}$\textbf{3.99}_{\pm{0.04}}$ & \cellcolor{magenta!15}$\textbf{3.59}_{\pm{0.06}}$ & \cellcolor{magenta!15}$\textbf{3.26}_{\pm{0.04}}$ & \cellcolor{magenta!15}$\textbf{92.73}_{\pm{0.01}}$ & \cellcolor{magenta!15}$\textbf{2.28}_{\pm{0.08}}$ & \cellcolor{magenta!15}$\textbf{3.61}_{\pm{0.03}}$ & \cellcolor{magenta!15}$\textbf{2.99}_{\pm{0.06}}$ & \cellcolor{magenta!15}$\textbf{2.91}_{\pm{0.05}}$ \\ \bottomrule
\end{tabular}}
\caption{\small Comparison of ProSE with different diffusion-based SE methods on VBD and TIMIT+MUSAN (synthetically noised datasets). ProSE outperforms our baselines by 0.5\% - 13\% with just 2 diffusion steps for inference.}
\label{tab:vbd_chime}
\end{table*}

\begin{align}
    q\left(\mathbf{z}^{\operatorname{S1}}_{1: T} \mid \mathbf{z}_0^{\operatorname{S1}}\right) 
    &= \prod_{t=1}^T q\left(\mathbf{z}^{\operatorname{S1}}_t \mid \mathbf{z}^{\operatorname{S1}}_{t-1}\right), \nonumber \\
    q\left(\mathbf{z}^{\operatorname{S1}}_t \mid \mathbf{z}^{\operatorname{S1}}_{t-1}\right) 
    &= \mathcal{N}\left(\mathbf{z}_t ; \sqrt{1-\beta_t} \mathbf{z}^{\operatorname{S1}}_{t-1}, \beta_t \mathbf{I}\right)
\end{align}
where $t= 1, \cdots, T$; $\mathbf{z}^{\operatorname{S1}}_{t}$ represents the noisy features at the $t$-th step; $\beta_{1: T} \in(0,1)$ is the pre-set noise schedule and $\mathcal{N}$ denotes the Gaussian distribution. Through reparameterization~\cite{kingma2013auto}, we can rewrite this as:
\begin{align}
    q\left(\mathbf{z}^{\operatorname{S1}}_t \mid \mathbf{z}^{\operatorname{S1}}_0\right) 
    &= \mathcal{N}\left(\mathbf{z}^{\operatorname{S1}}_t ; \sqrt{\bar{\alpha}_t} \mathbf{z}^{\operatorname{S1}}_0, \left(1-\bar{\alpha}_t\right) \mathbf{I}\right), \nonumber \\
    \alpha &= 1-\beta_t, \nonumber \\
    \bar{\alpha}_t &= \prod_{i=1}^t \alpha_i
\end{align}
During training, for the \textit{\textbf{reverse diffusion process}}, we generate the prior feature by learning to remove the noise added in the forward diffusion process. For the step from $\mathbf{z}^{\operatorname{S1}}_t$ to $\mathbf{z}^{\operatorname{S1}}_{t-1}$, we use the posterior distribution as:
\begin{align}
    q\left(\mathbf{z}^{\operatorname{S1}}_{t-1} \mid \mathbf{z}^{\operatorname{S1}}_t, \mathbf{z}^{\operatorname{S1}}_0\right) 
    &= \mathcal{N}\left(\mathbf{z}^{\operatorname{S1}}_{t-1} ; \boldsymbol{\mu}_t\left(\mathbf{z}^{\operatorname{S1}}_t, \mathbf{z}^{\operatorname{S1}}_0\right), \right. \nonumber \\
    &\quad \left. \frac{1-\bar{\alpha}_{t-1}}{1-\bar{\alpha}_t} \beta_t \mathbf{I} \right)
    \label{eqtn:cond}
\end{align}
\begin{equation}
    \boldsymbol{\mu}_t\left(\mathbf{z}_t, \mathbf{z}^{\operatorname{S1}}_0\right)=\frac{1}{\sqrt{\alpha_t}}\left(\mathbf{z}^{\operatorname{S1}}_t-\frac{1-\alpha_t}{\sqrt{1-\bar{\alpha}_t}} \boldsymbol{\epsilon}\right)
\end{equation}
where $\epsilon$ represents the noise in $\mathbf{z}^{\operatorname{S1}}_t$. For LDMs, a neural network estimates the noise at each step. At this stage, we add another latent encoder $\operatorname{LE_{S2}}$, which is similar to $\operatorname{LE_{S1}}$ except for the number of kernel channels in the input convolution layer. We employ the latent encoder $\operatorname{LE_{S2}}$ to generate the prior $\mathbf{z}^{\operatorname{S2}} \in \mathbb{R}^{N \times C^{\prime}}$ from $\mathbf{X}_{noisy}$. We now condition the LDM on $\mathbf{z}^{\operatorname{S2}}$ to perform noise estimation, i.e., $\boldsymbol{\epsilon}_\theta\left(\mathbf{z}^{\operatorname{S1}}_t, \mathbf{z}^{\operatorname{S2}}, t\right)$.
\begin{align}
    \mathbf{z}^{\operatorname{S1}}_{t-1} 
    &= \frac{1}{\sqrt{\alpha_t}}\left(\boldsymbol{y}_t - \frac{1-\alpha_t}{\sqrt{1-\bar{\alpha}_t}} \boldsymbol{\epsilon}_\theta\left(\mathbf{z}^{\operatorname{S1}}_t, \mathbf{z}^{\operatorname{S2}}, t\right)\right) \nonumber \\
    &\hspace{3cm}+ \sqrt{1-\alpha_t} \boldsymbol{\epsilon}_t,
\end{align}
where $\boldsymbol{\epsilon}_t \sim \mathcal{N}(0, \mathbf{I})$. By iterative denosing ($T$ times), we can generate the predicted prior feature $\hat{\mathbf{z}} \in \mathbb{R}^{N \times C}$. $\hat{\mathbf{z}}$ is then used to guide the SE Transformer ($\mathbf{z}_1$ =  $\hat{\mathbf{z}}$).  With much smaller dimensions, $\hat{\mathbf{z}}$ can be predicted with only a few iterations, and thus, we train the LDM jointly with the SE Transformer. To train the diffusion model for accurate prior estimation, we optimize an additional diffusion loss $\mathcal{L}_{\text {diff }}$ between the diffusion estimated prior $\hat{\mathbf{z}}$ and the actual prior $\mathbf{z}^{\operatorname{S1}}$ as follows:
\begin{equation}
    \mathcal{L}_{\text {diff }}=\|\hat{\mathbf{z}}-\mathbf{z}^{\operatorname{S1}}\|_1, 
\end{equation}
This differs from optimizing the weighted variational bound adopted in a wealth of prior work for LDM training~\cite{ho2020denoising}. However, this approach solely optimizes the LDM and does not ensure alignment between the estimated and original priors, which is essential for ProSE. Our final training objective $\mathcal{L}_{\text {all }}$ for joint training can be defined as follows:
\begin{equation}
    \mathcal{L}_{\text {all }}=\mathcal{L}_{\text {SE}}+\mathcal{L}_{\text {diff }}
\end{equation}

\subsubsection{Inference}
\label{sec:inference}

For inference, given a noisy speech signal $\mathbf{S}_{noisy}$, we first obtain its latent encoding $\mathbf{X}_{noisy}$ and then compress it to a compact space $\mathbf{z}^{\operatorname{S2}} \in \mathbb{R}^{N \times C^{\prime}}$ using $\operatorname{LE_{S2}}$. We then generate the prior $\hat{\mathbf{z}}$ using the LDM. Precisely, we perform the reverse process $T$ times starting from a randomly sampled Gaussian Noise $\boldsymbol{\epsilon} \sim \mathcal{N}(0, \mathbf{I})$. The generation is conditioned on $\mathbf{z}^{\operatorname{S2}}$ (Equation~\ref{eqtn:cond}). The generated prior is now conditioned on the SE Transformer to guide it for enhancement. The SE Transformer takes $\mathbf{X}_{noisy}$ as input and predicts clean speech encoding $\hat{\mathbf{X}}_{clean}$. $\hat{\mathbf{X}}_{clean}$ is then passed through a VAE decoder followed by a HiFi-GAN vocoder to obtain the $\hat{\mathbf{S}}_{clean}$.
\vspace{-2mm}

\section{Experimental Setup}
\label{sec:experiments}
\vspace{-1mm}
\begin{figure*}[t]
    \centering
    \includegraphics[width=1.0\textwidth]{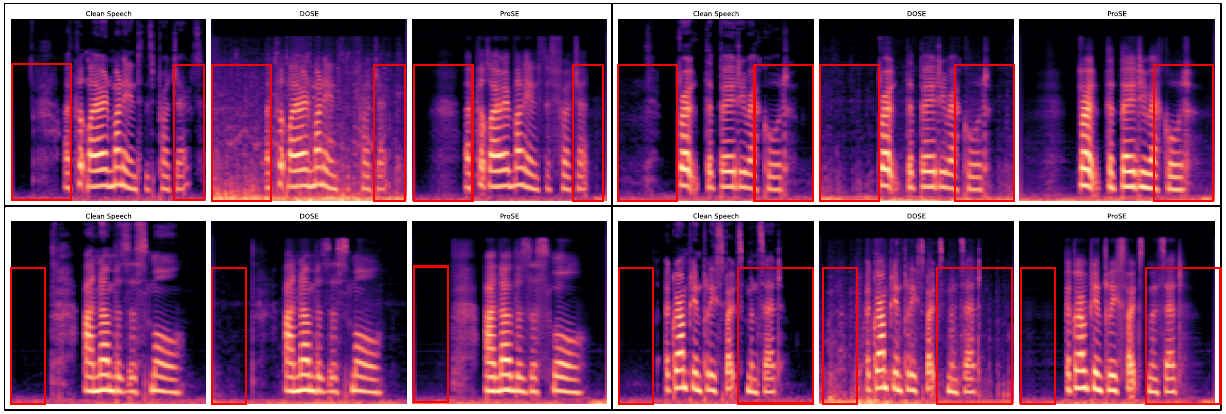}
    \caption{\small Comparison of mel-spectrograms generated by ProSE with DOSE (SOTA diffusion approach) and clean speech. ProSE generates enhanced speech that is closer to clean speech and does not add artifacts that are not present in the clean speech.}
    \label{fig:artifacts}
\end{figure*}
{\noindent \textbf{Datasets.}} Following a wealth of prior work in SE~\cite{lu2021study,lu2022conditional,tai2023revisiting,tai2023dose}, we use the VoiceBank-DEMAND (VBD) dataset to train and evaluate our models. The dataset comprises noisy speech recordings generated by mixing clean speech from the VoiceBank collection~\cite{veaux2013voice} with noise from the DEMAND dataset~\cite{thiemann2013diverse}. It has 30 distinct speakers, with 28 speakers allocated
for training and 2 for testing. Clean samples were mixed
with noise samples at four signal-to-noise ratios (SNRs) during training ([0, 5, 10, 15] dB) and testing ([2.5, 7.5, 12.5, 17.5] dB). The training dataset comprised a total of 11,572 utterances, while the testing set included 824 utterances. Additionally, to evaluate the generalization capabilities of our models trained on VBD, we also test them on CHiME-4~\cite{barker2015third}. CHiME-4 is sourced from the Wall Street Journal corpus with sentences spoken by talkers situated in challenging, noisy environments. In addition to VBD and CHiME-4, we also train and evaluate our models on TIMIT-MUSAN corpus. We synthesize this dataset by mixing clean speech from the TIMIT dataset~\cite{garofolo1993timit} with noise from the MUSAN dataset~\cite{snyder2015musan}. The mixing is done at SNRs ([-3, 0, 3, 5] dB).

\begin{table}[t]
\small
\centering
\resizebox{\columnwidth}{!}{
\begin{tabular}{lccccc}
\toprule
\textbf{Method} & \textbf{STOI (\%)}$\uparrow$ & \textbf{PESQ}$\uparrow$ & \textbf{CSIG}$\uparrow$ & \textbf{CBAK}$\uparrow$ & \textbf{COVL}$\uparrow$ \\ \midrule
\cellcolor{gray!15} Unprocessed & \cellcolor{gray!15} 71.50      & \cellcolor{gray!15} 1.21     & \cellcolor{gray!15} 2.18       & \cellcolor{gray!15} 1.97     & \cellcolor{gray!15} 1.62  \\ 
\noalign{\vskip 1mm}
\cdashline{1-6}
\noalign{\vskip 1mm}
Diffwave & 72.35     & 1.20      & 2.23       & 1.97     & 1.65  \\
DiffuSE  & 83.66  & 1.57   & 2.89   & 2.18     & \underline{2.18} \\
CDiffuSE & 82.75     & \underline{1.57} & 2.87       & 2.13     & 2.17 \\
SGMSE  & 84.41   & 1.56   & \textbf{2.92}  & \underline{2.18}  & 2.16 \\
DR-DiffuSE   & 78.03     & 1.31     & 2.41       & 2.07     & 1.82 \\
DOSE   & \underline{85.42}& 1.50      & 2.70       & 2.14     & 2.07 \\ 
\noalign{\vskip 1mm}
\cdashline{1-6}
\noalign{\vskip 1mm}
SE Transformer \textit{(ours)}  & $83.90_{\pm{0.03}}$           & $1.39_{\pm{0.05}}$           & $2.84_{\pm{0.03}}$             & $2.09_{\pm{0.04}}$           & $1.99_{\pm{0.01}}$ \\ \
\textbf{ProSE} \textit{(ours)}   & \cellcolor{magenta!15}$\textbf{88.18}_{\pm{0.02}}$ & \cellcolor{magenta!15}$\textbf{1.73}_{\pm{0.04}}$ & \cellcolor{magenta!15}$\underline{2.89}_{\pm{0.01}}$ & \cellcolor{magenta!15}$\textbf{2.25}_{\pm{0.07}}$ & \cellcolor{magenta!15}$\textbf{2.30}_{\pm{0.08}}$ \\
\bottomrule
\end{tabular}}
\caption{\small Comparison of ProSE with different diffusion-based SE methods on CHiME-4 (real-world dataset). ProSE outperforms our baselines by 0.7\% - 44.1\%.}
\label{tab:table2}
\end{table}
{\noindent \textbf{Evaluation metrics.}} We evaluate ProSE on the common evaluation metrics: perceptual evaluation of speech quality (PESQ)~\cite{rix2001perceptual}, short-time objective intelligibility (STOI)~\cite{}, segmental signal-to-noise ratio (SSNR)~\cite{taal2010short}, mean opinion score (MOS) prediction of the speech signal distortion (CSIG)~\cite{hu2007evaluation}, the MOS prediction of the intrusiveness of background noise (CBAK)~\cite{hu2007evaluation} and the MOS prediction of the overall effect (COVL)~\cite{hu2007evaluation}.

{\noindent \textbf{Baselines.}} We compare ProSE with recent open-sourced diffusion-based SE methods DiffuSE~\cite{lu2021study}, CDiffuSE~\cite{lu2022conditional}, SGMSE~\cite{richter2023speech}, DR-DiffuSE~\cite{tai2023revisiting} and DOSE~\cite{tai2023dose}. Details about baselines can be found in Appendix~\ref{sec:baseline}.

{\noindent\textbf{Hyper-parameters.}} We process speech at 16kHz. For mel-spectrogram conversion, we set the window, FFT, and hop sizes to 1024, 1024, and 160, respectively, with fmin and fmax set to 0 and 8000. This results in F=128, and we slice larger audios and pad smaller ones to a fixed duration of 2 seconds, which results in T=200. We employ $r$=4 for VAE. We train ProSE with an Adam optimizer~\cite{kingma2014adam} with $\beta_1$=0.9 and $\beta_2$=0.99. For stage one, the total training iterations are 300K. The initial learning rate is set as $2\times10^{-4}$ and gradually reduced to $1\times10^{-6}$ with the cosine annealing. For stage two, we adopt the same training settings as in stage one. For LDM, we use $T$ = 2 for training with the variance hyperparameters $\beta_{1:T}$ constants increasing linearly from $\beta_1$=0.1 to $\beta_T$ =0.99. We also employ $T$ = 2 during inference. We train both stages with a batch size of 8 on 4 A100 GPUs. We set N=16 and $C^{\prime}$=256 as the LE output dimensions for our experiments. For the first 4 levels of the SE Transformer, the number of Transformer blocks is [3,5,5,6], the number of channels is [48,96,192,384], and the attention heads are [1,2,4,8]. 
\vspace{-2mm}

\section{Results}
\label{sec:results}
\vspace{-1mm}

{\noindent \textbf{Quantitative Results.}} Table~\ref{tab:vbd_chime} (left) compares ProSE with other diffusion-based SE methods in the literature when trained and inferred on the VBD dataset (matched scenario). Table~\ref{tab:table2} compares ProSE with other diffusion-based SE methods in the literature when trained on VBD and inferred on the CHiME-4 dataset (unmatched scenario). We show that: \textbf{(1)} diffusion-based methods offer superior generalizability compared to deterministic approaches, \textbf{(2)} techniques that incorporate specific condition-injecting strategies, such as DiffuSE, CDiffuSE, and SGMSE, demonstrate robust generalizability, though they slightly underperform deterministic mapping-based methods in matched scenarios, and \textbf{(3)} ProSE outperforms other diffusion-based SE models in both matched and unmatched scenarios and does so with fewer computational steps. Table~\ref{tab:vbd_chime} (right) further compares ProSE when trained and inferred on TIMIT+MUSAN. All 3 points above hold, and ProSE outperforms our baselines on all metrics by significant margins. All the results shown in both the tables are averaged across 3 runs. Additionally, as an ablation of ProSE, we show the importance of the LDM prior to the SE Transformer. For all tables, the SE Transformer rows show the performance of the U-Net transformer without conditioning of the LDM prior, i.e., we remove the conditioning block and train the model for only a single stage with a regression objective. As we see, without the LDM prior, performance decreases substantially across all metrics and all datasets.
\vspace{1mm}
\begin{figure*}[t]
    \centering
    \includegraphics[trim=0 1cm 0 0, width=0.95\textwidth]{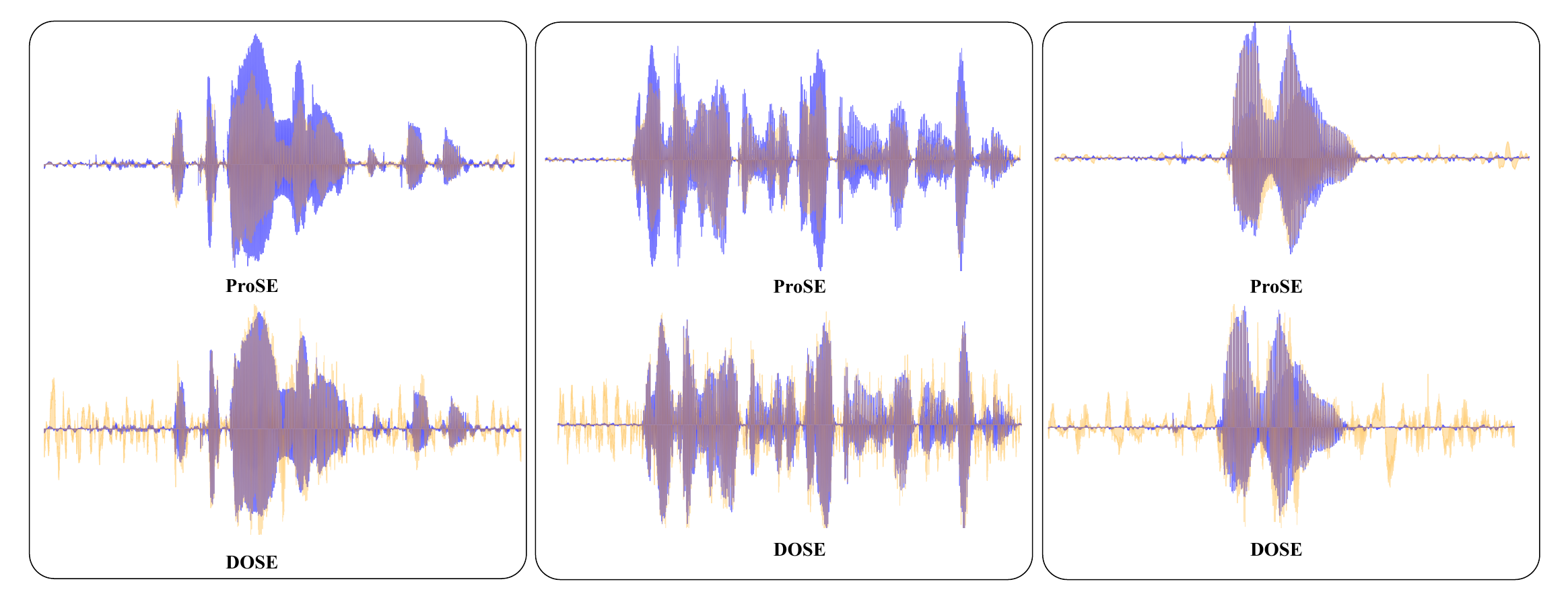}
    \caption{\small Comparison of waveforms generated by ProSE with DOSE (SOTA diffusion approach) and clean speech. The blue waveform is the original clean speech, and the yellow waveform is the enhanced speech generated by the model. We show that while DOSE generates additional details that are misaligned with the original speech, ProSE does not face this issue.}
    \label{fig:artifacts_wav}
\end{figure*}

{\noindent \textbf{Qualitative Results.}} In addition to quantitative metrics, we employ two varieties of Mean Opinion Score (MOS) tests to assess the quality of synthesized speech via human evaluation. These 2 metrics include Naturalness and Consistency, as proposed by Tai \textit{et. al}~\cite{tai2023dose}. Table~\ref{tab:mos_result} shows that ProSE outperforms all baseline models in generating natural-sounding clean speech and matching the quality and content of the reference gold-quality clean speech. 

\vspace{-3pt}
\section{Discussion}
\label{sec:discuss}
\vspace{-1mm}

\begin{figure}[t]
     \centering
         \includegraphics[trim= 0cm 0.5cm 0cm 0cm, width=0.5\textwidth]{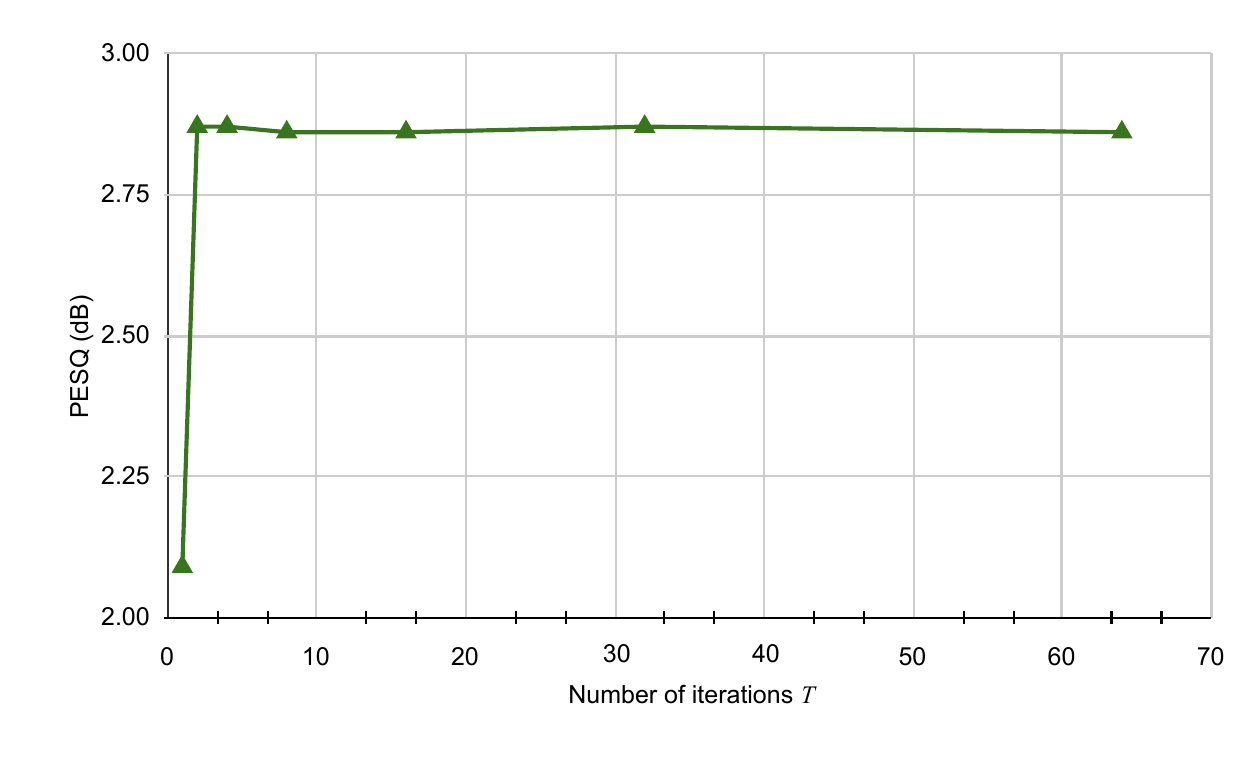}
     \caption{\small Effect of the number of inference timesteps $T$ on PESQ for ProSE, trained and evaluated on VBD.}
        \label{fig:effect_t}
        \vspace{-1em}
\end{figure}

{\noindent \textbf{Effect of the number of iterations $T$ in LDM.}} Fig.~\ref{fig:effect_t} shows the effect of inference timesteps $T$ on the PESQ score for ProSE. The results shown are for ProSE trained and evaluated on VBD. As we can see, ProSE achieves optimal performance on only $T$=2 iterations (by default, we use the same $T$ for training and inference). We attribute this characteristic to the way we integrate DMs into the ProSE architecture: precisely, we only apply DMs in a compact latent space, and thus, the model does not need too many iterations to add details to the output, unlike speech synthesis. Additionally, the latent generated, which is used as a prior for conditioning, only adds details for enhancement. Thus, only a few iterations can lead to a stable prior. This also makes ProSE much more efficient than prior work, which employs DMs to generate the entire enhanced speech.

{\noindent \textbf{Waveform and Mel-spectrogram Analysis.}} Figures \ref{fig:artifacts_wav} and \ref{fig:artifacts} display and compare waveforms and mel-spectrograms, respectively. We present a three-way analysis involving ground-truth clean speech, speech enhanced by ProSE, and speech enhanced by a SOTA diffusion-based model, DOSE. Notably, DOSE tends to introduce extraneous misaligned details, a problem not observed with ProSE. This distinction arises because ProSE utilizes DMs solely to generate a prior, with the actual enhancement performed by a regression model. This strategy prevents the misalignment issues commonly seen when the iterative denoising process is used to generate enhanced speech over multiple steps directly.

\begin{table}[t]
\centering
\resizebox{0.6\columnwidth}{!}{
\begin{tabular}{lcc}
\toprule
\textbf{Method}    & \textbf{FLOPs} (G)$\downarrow$ & \textbf{PESQ} (dB)$\uparrow$ \\ \midrule
\textbf{DiffWave}     & 180.25      & 2.50 \\
\textbf{DiffuSE}      & 197.65      & 2.37 \\
\textbf{CDiffuSE}     & 227.40      & 2.41 \\
\textbf{SGMSE}        & \textbf{145.47} & 2.34 \\
\textbf{DR-DiffuSE}   & 289.58      & 2.48 \\
\textbf{DOSE}         & 174.25      & \underline{2.54} \\
\textbf{ProSE}        & \underline{150.78} & \textbf{2.87} \\
\bottomrule
\end{tabular}}
\caption{\small Model complexity comparison of ProSE.}
\label{tab:params_table_prose}
\vspace{-2em}
\end{table}

{\noindent \textbf{Computational Efficiency.}} Table~\ref{tab:params_table_prose} compares the computational efficiency of ProSE with other baselines from the literature. All experiments are conducted on the VBD dataset and we report inference FLOPs. ProSE is more computationally efficient than most baselines while achieving SOTA performance in SE. We attribute this to applying LDM in the latent space, which allows us to achieve better performance with fewer LDM iterations, as opposed to all our baselines where DM is employed to generate enhanced speech directly.


\section{Conclusion}
\label{sec:conclusion}

In this paper, we propose ProSE, a novel method for speech enhancement with diffusion models. Specifically, ProSE makes a latent diffusion model generate a prior. This prior is then conditioned on a U-Net style regression-based transformer network where the prior guides the network for speech enhancement. The regression-based method preserves the general distribution, while the prior feature generated by the diffusion model enhances the details of the enhanced speech. ProSE achieves SOTA performance on standard benchmarks in matched and unmatched scenarios.

\section{Acknowledgment}
This project is supported in part by NSF\#1910940.

\section*{Limitations and Future Work}

As part of future work, we would like to address the current limitations of ProSE, including: 

\begin{enumerate}
    \item The requirement of VAE and HiFi-GAN modules: Though these modules are standard practice in speech and audio systems, loss of information in these modules is generally propagated to the final performance, adding to the computational overhead. Therefore, a transformer architecture with raw audio, input, and output would help overcome this problem. 
    \item The requirement of 2 stages of training: Our proposed algorithm is built on fundamentals that require two stage training, whereby in the first stage we learn a prior and in the second stage we learn the DDPM models. As part of future work we would like to work towards simplifying ProSE's training pipeline.
\end{enumerate}

\bibliography{anthology,custom}

\appendix

\section{Appendix}
\label{sec:additional}

In the Appendix, we provide:
\begin{enumerate}
    \item Section~\ref{sec:vae_ad}: Variational Auto Encoder (VAE) Architecture Details
    \item Section~\ref{sec:vae_training}: Variational Auto Encoder Training Details
    \item Section~\ref{sec:hifi}: HiFi-GAN Training Details
    \item Section~\ref{sec:license}: Dataset Licences
    \item Section~\ref{sec:baseline}: Baseline Details
    \item Section~\ref{sec:mos_test}: MOS Test
    \item Section~\ref{sec:broader}: Broader Impact
\end{enumerate}

\section{Variational Auto Encoder (VAE) Architecture Details}
\label{sec:vae_ad}
We compress the mel-spectrogram $\ell \in \mathbb{R}^{\text{T} \times \text{F}}$ into a small continuous space $\mathbf{X} \in \mathbb{R}^{C \times \frac{\text{T}}{r} \times \frac{\text{F}}{r}}$ with a convolutional VAE, where $T$ and $F$ represent the time and frequency dimensions respectively, $C$ is the channel number of the latent encoding, and $r$ is the compression level (downsampling ratio) of the latent space. Both the encoder $\mathcal{E}(\cdot)$ and the decoder $\mathcal{D}(\cdot)$ are composed of stacked convolutional modules. This allows the VAE encoder to preserve the spatial correspondence between the mel-spectrogram and the latent space, as shown in Figure~\ref{fig:main_diag}. Each module consists of ResNet blocks~\cite{he2016deep}, which include convolutional layers and residual connections. The encoding $\mathbf{X}$ is evenly split into two parts, $\mathbf{X}_\mu$ and $\mathbf{X}_\sigma$, with shape C, $\frac{\text{T}}{r}, \frac{\text{F}}{r}$, representing the mean and variance of the VAE latent space. The input of the decoder is a stochastic encoding $\hat{z} = z_\mu + z_\sigma \cdot \epsilon$, where $\epsilon \sim \mathcal{N}(0, I)$. During generation, the decoder reconstructs the mel-spectrogram from the generated latent representations.

We employ three loss functions in our training objective: the mel-spectrogram reconstruction loss, adversarial losses, and a Gaussian constraint loss. The reconstruction loss calculates the mean absolute error between the input sample $\mathbf{X} \in \mathbb{R}^{\text{T} \times \text{F}}$ and the reconstructed mel-spectrogram $\hat{\ell} \in \mathbb{R}^{T \times F}$. The adversarial losses are used to enhance the reconstruction quality. Specifically, we adopt the PatchGAN~\cite{isola2017image} as our discriminator, which divides the input image into small patches and predicts whether each patch is real or fake by outputting a matrix of logits. The PatchGAN discriminator is trained to maximize the logits for correctly identifying real patches while minimizing the logits for incorrectly identifying fake patches. We also apply the Gaussian constraint on the latent space of the VAE. By enforcing a Gaussian constraint on the latent space, the VAE is encouraged to learn a continuous, structured latent space rather than a disorganized one. This helps the VAE to better capture the underlying structure of the data, leading to more stable and accurate reconstructions~\cite{kingma2013auto}.

\section{Variational Auto Encoder Training Details}
\label{sec:vae_training}
We train our VAE using the Adam optimizer~\cite{kingma2014adam} with a learning rate of $4.5 \times 10^{-6}$ and a batch size of six. The speech data we use for training is VoiceBank-DEMAND. We perform experiments with three compression-level settings r=4, 8, 16, for which the latent channels are C=8, 16, and 32, respectively. VAEs in all three settings are trained with at least 1.5M steps on a single NVIDIA RTX 3090 GPU. To stabilize training, we do not apply the adversarial loss in the first 50K training steps. We use the mixup~\cite{zhang2018mixup} strategy for data augmentation.

\section{HiFi-GAN Training Details}
\label{sec:hifi}

In this study, we utilize HiFi-GAN~\cite{kong2020hifi}, renowned for its effectiveness in speech waveform generation, as our vocoder. HiFi-GAN employs two sets of discriminators, a multi-period discriminator, and a multi-scale discriminator, to enhance perceptual quality. We train the vocoder on the VoiceBank-DEMAND dataset to synthesize audio waveforms. For input samples with a sampling rate of 16,000Hz, we extract 128-band mel-spectrograms and adhere to the default HiFi-GAN V1 settings. The window, FFT, and hop sizes are set to 1024, 1024, and 160, respectively, with fmin and fmax set to 0 and 8000. We employ the AdamW optimizer with $\beta_1$=0.8 and $\beta_2$=0.99. The learning rate is initialized at $2 \times 10^-4$ and decays at a rate of 0.999. Using a batch size of 96, we train the model on six NVIDIA 3090 GPUs.

\section{Dataset Licenses}
\label{sec:license}
\textbf{VoiceBank-DEMAND~\footnote{\url{http://homepages.inf.ed.ac.uk/jyamagis/release/VCTK-Corpus.tar.gz}}}: Licensed under CC BY 4.0.\\
\textbf{CHiME-4 (WSJ)~\footnote{\url{https://www.chimechallenge.org/challenges/chime4/index}}}: Licensed under LDC User Agreement for Non-Members.\\
\textbf{TIMIT~\footnote{\url{https://catalog.ldc.upenn.edu/LDC93S1}}}: Licensed under LDC User Agreement for Non-Members.\\
\textbf{MUSAN~\footnote{\url{https://www.openslr.org/17/}}}: Licensed under CC BY 4.0.

\section{Baseline Details}
\label{sec:baseline}
{\noindent \textbf{DiffWave}}: DiffWave~\cite{kong2021diffwave}~\footnote{\url{https://github.com/lmnt-com/diffwave}} proposes a versatile non-autoregressive diffusion probabilistic model designed for conditional and unconditional audio synthesis. By converting white noise into structured waveforms through a Markov chain, DiffWave achieves high-fidelity audio generation across various tasks, including neural vocoding and class-conditional generation. The model optimizes a variant of the variational bound on data likelihood and demonstrates superior performance compared to autoregressive and GAN-based models, particularly in unconditional audio generation tasks, achieving high speech quality and synthesis speed. Licensed under Apache-2.0 license.

{\noindent \textbf{DiffuSE}}: DiffuSE~\cite{lu2021study}~\footnote{\url{https://github.com/neillu23/DiffuSE}} proposes a speech enhancement model based on diffusion models. DiffuSE uses a novel supportive reverse process to eliminate noise by combining noisy speech signals at each step, improving the quality of enhanced speech. The model is pretrained with clean Mel-spectral features and fine-tuned using noisy spectral features. Experimental results show that DiffuSE outperforms existing time-domain generative SE models, providing significant improvements in speech quality and robustness. Licensed under Apache-2.0 license.

{\noindent \textbf{CDiffuSE}}:
CDiffuSE~\cite{lu2022conditional}~\footnote{\url{https://github.com/neillu23/CDiffuSE}} proposes a conditional diffusion probabilistic model for speech enhancement. By incorporating noisy speech signals into both diffusion and reverse processes, CDiffuSE adapts better to non-Gaussian noise, improving the quality of enhanced speech. Experimental results show CDiffuSE outperforms traditional generative models and demonstrates strong generalizability  to unseen noise conditions. This method significantly enhances speech clarity and robustness compared to state-of-the-art approaches. Licensed under Apache-2.0 license.

{\noindent \textbf{SGMSE}}: SGMSE~\cite{richter2023speech}~\footnote{\url{https://github.com/sp-uhh/sgmse}}  presents a diffusion-based generative model for speech enhancement and dereverberation, building upon a stochastic differential equation framework. Unlike traditional conditional generation tasks, the model begins the reverse process from a mixture of noisy speech and Gaussian noise, aligning it with the forward process that transitions from clean to noisy speech. By adapting the network architecture and leveraging only 30 diffusion steps, the model significantly improves speech quality and generalization across different datasets. Experimental results show that this method outperforms recent discriminative models and excels in real-world noisy conditions, enhancing both additive noise removal and dereverberation. Licensed under MIT license.

{\noindent \textbf{DR-DiffuSE}}: DR-DiffuSE~\cite{tai2023revisiting}~\footnote{\url{https://github.com/judiebig/DR-DiffuSE}} addresses challenges in applying DDPMs to speech enhancement by introducing techniques to tackle condition collapse and improve inference efficiency. The method combines an auxiliary conditional generation network, a dual-path parallel network, and a fast sampling technique, followed by a refinement network to calibrate the generated speech. Experimental results demonstrate that DR-DiffuSE significantly improves speech quality and robustness, outperforming existing DDPM-based and other generative SE models.

{\noindent \textbf{DOSE}}: DOSE~\cite{tai2023dose}~\footnote{\url{https://github.com/ICDM-UESTC/DOSE}} introduces a novel speech enhancement method utilizing diffusion dropout with an adaptive prior. DOSE addresses the challenge of incorporating condition information in diffusion probabilistic models by employing dropout during training to prioritize condition factors and using an adaptive prior to guide the sampling process. This approach significantly enhances the quality and stability of generated speech, improving consistency with condition factors and inference efficiency. Experimental results demonstrate that DOSE surpasses existing diffusion-based and deterministic methods in terms of speech quality and robustness.

\section{MOS Test}
\label{sec:mos_test}

\noindent We perform two types of Mean Opinion Score (MOS) tests to assess the quality of generated audio through human evaluation.

Naturalness: For this test, we ask the raters to evaluate the audio quality and naturalness while ignoring differences in style (timbre, emotion, and prosody). The raters listen to and rate the samples, scoring the naturalness on a 1-5 Likert scale.

Consistency: For this test, we instruct the raters to focus on how similar the generated speech is to the reference in terms of content, timbre, emotion, and prosody, while ignoring audio quality. This is slightly different from the original Similarity Mean Opinion Score (SMOS) test. In SMOS tests, each generated utterance is paired with a ground truth utterance to see how well the generated speech matches the target speaker. The raters listen to and rate the samples, scoring the consistency on a 1-5 Likert scale.

\begin{figure*}
    \centering
    \includegraphics[width=\textwidth]{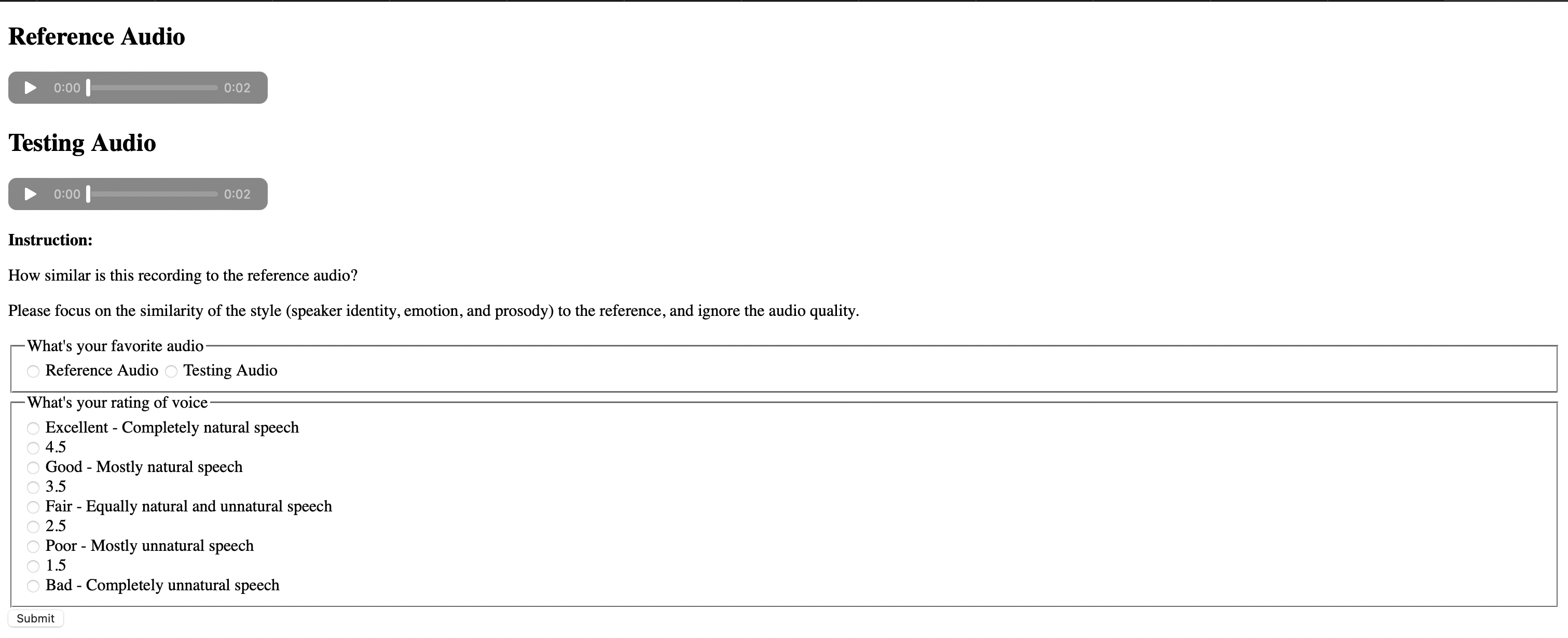}
    \caption{Similarity MOS Test Application UI}
    \label{fig:mos_test_sim}
\end{figure*}

\begin{figure*}
    \centering
    \includegraphics[width=\textwidth]{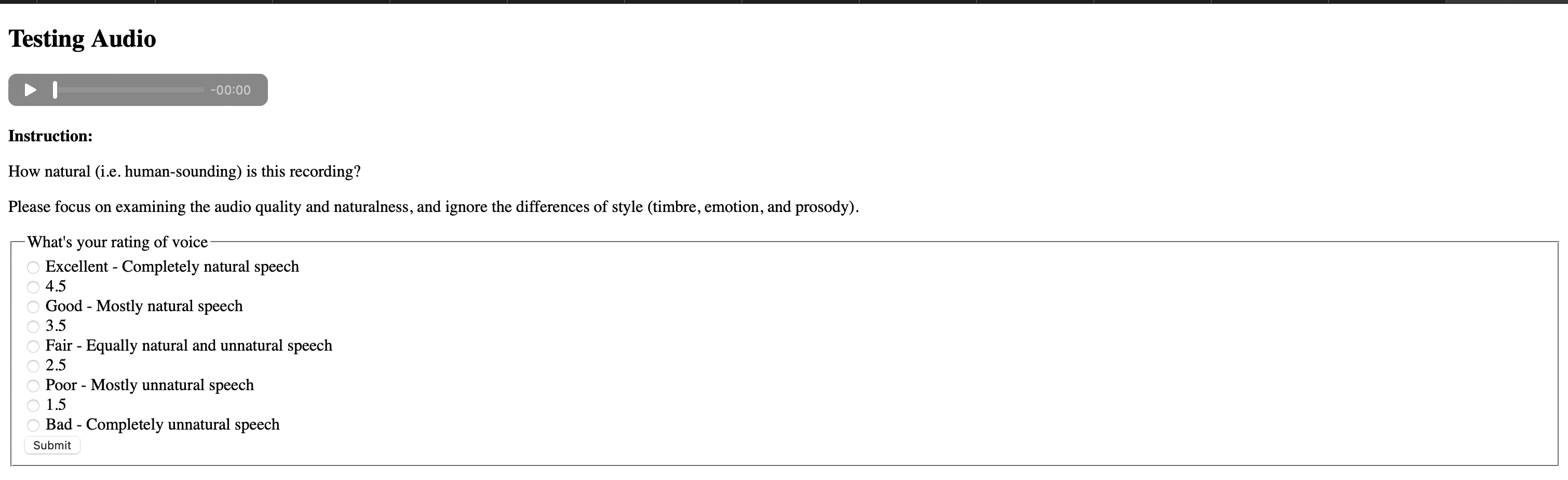}
    \caption{MOS Test Application UI}
    \label{fig:mos_test}
\end{figure*}

We conduct these subjective evaluations with the help of 20 volunteers, and the instructions for the testers are shown in Figure \ref{fig:mos_test} and Figure \ref{fig:mos_test_sim}. The MOS results with 95\% confidence intervals are shown in Table \ref{tab:mos_result}. Based on our test, we find: (1) Our method outperforms all baselines, showing its strong capability in producing natural-sounding speech; (2) Our model generates speech that is consistent with the reference speech, aligning with our design goals.
\label{sec:mos_test}

\begin{table*}
\centering
\resizebox{\textwidth}{!}{%
\begin{tabular}{lcccccc}
\toprule
\textbf{Method}        & {Scenarios} & {MOS$(\uparrow)$}  & {Similarity MOS$(\uparrow)$} & {Scenarios} & {MOS$(\uparrow)$}  & {Similarity MOS$(\uparrow)$} \\ \midrule 
DiffWave       &   {\multirow{8}{*}{Matched}}       & 3.65 & 3.27           &    \multirow{8}{*}{Mismatched}       & 3.20 & 3.05           \\
DiffuSE        &           & 3.75 & 3.43           &           & 2.75 & 1.63           \\
CDiffuSE       &           & 3.45 & 3.31           &           & 2.80  & 1.97           \\
SGMSE          &           & 3.60  & 3.45           &           & 3.10  & 2.21           \\
DR-DiffuSE     &           & 3.55 & 3.47           &           & 3.05 & 2.03           \\
DOSE           &           & 3.70  & 3.62           &           & 2.95 & 2.28           \\
SE Transformer &           & 3.85 & 3.71           &           & 3.35 & 3.17           \\
ProSE          &           & \textbf{4.45} & \textbf{4.08}           &           & \textbf{3.45} & \textbf{3.23}     \\ \hline   
\end{tabular}}
\caption{\small MOS Test Result}
\label{tab:mos_result}
\end{table*}

\section{Broader Impact}
\label{sec:broader}

We introduce ProSE, a new approach to speech enhancement (SE) that significantly improves the clarity and quality of speech in noisy environments. By integrating denoising diffusion probabilistic models with Transformer-based regression, ProSE offers a more efficient solution that requires less computational power compared to traditional methods. This advancement is particularly important for real-time applications like voice assistants, telecommunication, and hearing aids, where immediate and clear speech output is crucial. Moreover, the reduced computational demand makes this technology more accessible and practical for implementation in various devices and systems. ProSE is taking a step forward in enhancing the user experience and broadening the accessibility of clear communication technology.

\vspace{0.5em}
{\noindent \textbf{Potential Risks.}} Our institution’s Institutional Review Board (IRB) thoroughly reviewed and approved the human study presented in the paper, ensuring that ethical guidelines and safety measures were adhered to throughout the research process.

\end{document}